\documentclass[conference]{IEEEtran}
\IEEEoverridecommandlockouts
\usepackage{cite}
\usepackage{amsmath,amssymb,amsfonts}
\usepackage{algorithmic}
\usepackage{graphicx}
\usepackage{textcomp}
\usepackage{xcolor}
\def\BibTeX{{\rm B\kern-.05em{\sc i\kern-.025em b}\kern-.08em
    T\kern-.1667em\lower.7ex\hbox{E}\kern-.125emX}}

\usepackage{amsmath,amsfonts,bm}

\def\eqref#1{equation~\ref{#1}}
\def\1{\bm{1}}

\DeclareMathAlphabet{\mathsfit}{\encodingdefault}{\sfdefault}{m}{sl}
\SetMathAlphabet{\mathsfit}{bold}{\encodingdefault}{\sfdefault}{bx}{n}

\usepackage{hyperref}
\usepackage{url}
\usepackage{graphicx}
\usepackage{float}
\usepackage{graphicx}
\usepackage{graphbox}
\usepackage{caption}
\usepackage{subcaption}
\usepackage{tabularx}
\usepackage{booktabs}
\usepackage{enumitem}
\usepackage{adjustbox}
\usepackage{chngcntr}
\usepackage{multirow}
    
\begin{document}

\title{Glioblastoma Tumor Segmentation using an Ensemble of Vision Transformers}
\author{\IEEEauthorblockN{Huafeng Liu\IEEEauthorrefmark{1},
Benjamin Dowdell\IEEEauthorrefmark{1},
Todd Engelder\IEEEauthorrefmark{1},
Zarah Pulmano\IEEEauthorrefmark{1},
Nicolas Osa\IEEEauthorrefmark{1}, and
Arko Barman\IEEEauthorrefmark{2}}
\IEEEauthorblockA{\IEEEauthorrefmark{1}Department of Computer Science, Rice University}
\IEEEauthorblockA{\IEEEauthorrefmark{2}Data to Knowledge Lab \& Department of Electrical and Computer Engineering, Rice University\\
Email: arko.barman@rice.edu}}

% \author{\IEEEauthorblockN{Huafeng Liu}
% \IEEEauthorblockA{\textit{Rice University, USA} \\
% Huafeng.Liu@rice.edu}
% \and
% \IEEEauthorblockN{Benjamin Dowdell}
% \IEEEauthorblockA{\textit{Rice University, USA} \\
% ben.dowdell@rice.edu}
% \and
% \IEEEauthorblockN{Todd Engelder}
% \IEEEauthorblockA{\textit{Rice University, USA} \\
% Todd.Engelder@rice.edu}

% \and
% \IEEEauthorblockN{Zarah Pulmano}
% \IEEEauthorblockA{\textit{Rice University, USA} \\
% kp52@rice.edu }
% \and
% \IEEEauthorblockN{Nicolas Osa}
% \IEEEauthorblockA{\textit{Rice University, USA} \\
% nao5@rice.edu}
% \and
% \IEEEauthorblockN{Arko Barman}
% \IEEEauthorblockA{\textit{Rice University, USA} \\
% arko.barman@rice.edu}
% }

\maketitle

\begin{abstract}
Glioblastoma is one of the most aggressive and deadliest types of brain cancer, with low survival rates compared to other types of cancer. Analysis of Magnetic Resonance Imaging (MRI) scans is one of the most effective methods for the diagnosis and treatment of brain cancers such as glioblastoma. Accurate tumor segmentation in MRI images is often required for treatment planning and risk assessment of treatment methods. Here, we propose a novel pipeline, Brain Radiology Aided by Intelligent Neural NETworks (BRAINNET), which leverages MaskFormer, a vision transformer model, and generates robust tumor segmentation maks. We use an ensemble of nine predictions from three models separately trained on each of the three orthogonal 2D slice directions (axial, sagittal, and coronal) of a 3D brain MRI volume. We train and test our models on the publicly available UPenn-GBM dataset, consisting of 3D multi-parametric MRI (mpMRI) scans from 611 subjects. Using Dice coefficient (DC) and 95\% Hausdorff distance (HD) for evaluation, our models achieved state-of-the-art results in segmenting all three different tumor regions -- tumor core (DC = 0.894, HD = 2.308), whole tumor (DC = 0.891, HD = 3.552), and enhancing tumor (DC = 0.812, HD = 1.608).
\end{abstract}


\section{Introduction}

\subsection{Background}

Glioblastoma (GBM) is one of the most aggressive and deadliest types of brain cancer, accounting for 49.1\% of all primary malignant cancers, and GBM-positive patients currently have a 6.8\% five-year survival rate \cite{10.1093/neuonc/noab200}. It can present as a heterogeneous tumor with several regions, including the necrotic tumor core (NCR), peritumoral edematous tissue (ED), and enhancing tumor (ET)\cite{BRATS2018}. Treatment options include either neurosurgery, radiation therapy, chemotherapy, immunotherapy, or targeted therapy\cite{https://doi.org/10.3322/caac.21613}. Treatment planning and risk assessment of treatment strategies greatly benefit from the segmentation of the tumor from imaging, which is thus a crucial task.

Magnetic resonance imaging (MRI) is routinely used in the diagnosis and treatment planning of GBM because it is non-invasive and can resolve subtle contrasts within soft tissue. MRI exams can include multiple modalities, depending on recording parameters and whether a contrast agent is used, resulting in multiple 3-dimensional volumes for a single patient\cite{BRATS2018}. The diagnosis and treatment planning of cancer using MRI requires a radiologist to review and interpret a large volume of 2D slices per patient. The current demand requires a radiologist to review a new 2D MRI image every 3 to 4 seconds continuously for 8 hours across 255 working days a year \cite{MCDONALD20151191}, also resulting in radiologist fatigue, thus impacting diagnostic efficacy and efficiency\cite{ZHAN2021424}. 

Artificial Intelligence (AI) assisted medical image analysis for diagnosis and treatment has been widely reported in recent years\cite{medicalAI, SelvathiPoornila2017, Zhang_Guo2018}. In this study, we introduce a neural network-based glioblastoma tumor segmentation model, \emph{Brain Radiology Aided by Intelligent Neural NETworks} (\emph{BRAINNET}), that leverages a vision transformer model called MaskFormer\cite{Maskformer}, along with an ensemble strategy using separate models trained on 2D axial, sagittal, and coronal slices of 3D multi-parametric MRI (mpMRI) images of the brain. We also apply transfer learning to simplify the training process by fine-tuning a pre-trained MaskFormer model to segment the GBM tumor. BRAINNET can be applied to reduce the time it takes for a radiologist to review MRI scans by reviewing the segmentation masks from the model side-by-side with the MRI image.

\subsection{Literature review}

Neural networks have improved both the efficiency and accuracy of medical image segmentation substantially. Several novel approaches, such as UNet \cite{UNet}, Region-based Convolutional Neural Network (R-CNN) \cite{Olmez2020-ta}, Mask R-CNN \cite{Mask_RCNN_tumor}, and state-of-the-art vision transformer models\cite{TransBTS} have been applied for segmentation tasks involving radiological images. For segmentation, MaskFormer \cite{Maskformer}, a vision transformer model, unifies the tasks of detection and segmentation in a single end-to-end trainable framework that can be used for instance-level segmentation, demonstrating strong performance on various instance segmentation benchmarks.

Starting in 2012, the International Brain Tumor Segmentation (BraTS) Challenge has been an annual competition for developing state-of-the-art machine learning models for brain tumor segmentation \cite{Menze2015-dg}. One of the earliest neural network-based architectures applied to this dataset used a multi-scale 3D Convolutional Neural Network (CNN) with a fully connected Conditional Random Field to remove false positives\cite{Kamnitsas_BraTS2015}. Another architecture trained a 3D auto-encoder with skip connections for brain tumor segmentation achieving state-of-the-art in the BraTS 2018 challenge\cite{3D_ae_tumor}. UNet-based architectures with sophisticated strategies, such as cascading, region-based training, and data augmentation, for brain tumor segmentation have also been proposed\cite{UNet_tumor, Jiang_BraTS2019, Isensee_BraTS2020, Luu_BraTS2021}. Models have been developed for uncertainty estimation and survival prediction based on segmentation along with the use of clinical data\cite{Jiang_BraTS2019, McKinley_BraTS2020}. In recent years, vision transformers have also been applied to obtain state-of-the-art performance in GBM tumor segmentation using the BraTS dataset\cite{TransBTS}.

\subsection{Our Contributions}

In our work, we develop and test BRAINNET, a pipeline that uses MaskFormer models and an ensemble approach to accomplish automated segmentation of GBM tumors. Our contributions include: (i) the development of an end-to-end pipeline for GBM segmentation using Maskformer models for mpMRI images; (ii) leveraging an ensemble strategy that combines the results from separate models for axial, sagittal, and coronal slices; (iii) achieving state-of-the-art performance with relatively low computational and memory requirements.

\section{Methods}
\subsection{Maskformer}

\begin{figure}[t]
\begin{center}
\includegraphics[width=\linewidth]{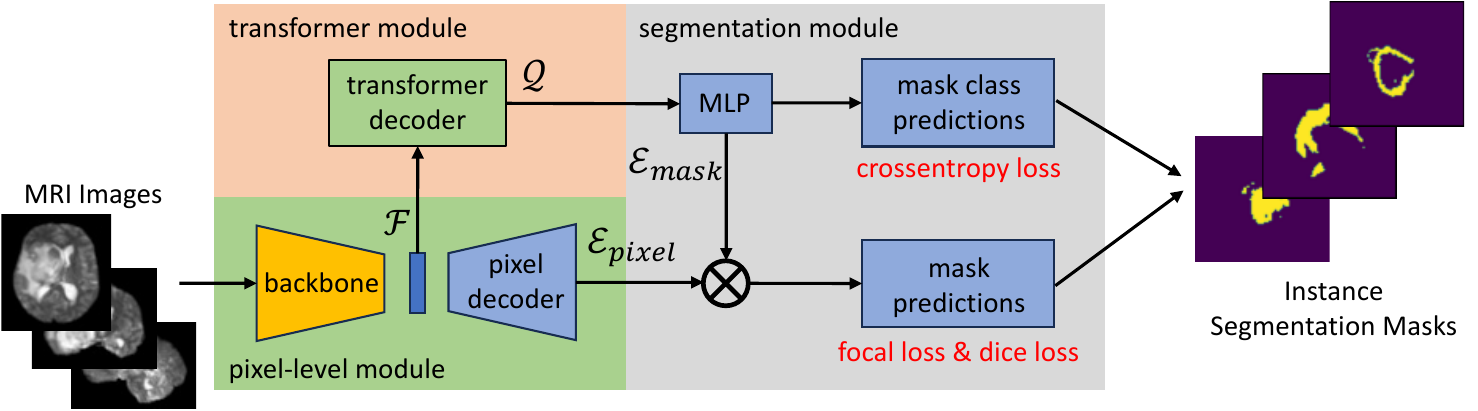}
\end{center}
\caption{MaskFormer model used in BRAINNET.}
\label{fig:Maskformer}
\end{figure}

We adapt the state-of-the-art MaskFormer model \cite{Maskformer}, a vision transformer model, for our proposed BRAINNET pipeline due to its simplicity, flexibility, and performance on general 2D image segmentation tasks. MaskFormer has an intuitive architecture with a flexible choice of the backbone (\autoref{fig:Maskformer}). It has three separate modules:
\begin{itemize}
    \item a pixel-level module that extracts image features, $\mathcal{F}$ using a backbone network and subsequently a pixel decoder network for upsampling $\mathcal{F}$ and extracting per-pixel embeddings, $\mathcal{E}_{\text{pixel}}$;
    \item a transformer module that feeds $\mathcal{F}$ into a transformer decoder to generate per-segment embeddings, $\mathcal{Q}$; and
    \item a segmentation module that generates $N$ class predictions along with $N$ corresponding mask embeddings, $\mathcal{E}_{\text{mask}}$ by feeding $\mathcal{Q}$ through a multilayer perceptron (MLP). The final mask predictions are generated by taking the dot product of $\mathcal{E}_{\text{pixel}}$ with $\mathcal{E}_{\text{mask}}$ and using a sigmoid activation.
\end{itemize}

In our experiments, we used a pre-trained MaskFormer\cite{Maskformer} with shifted windows (SWIN) transformer as the backbone\cite{Swin}, which was trained on the ADE20K, a scene segmentation benchmark dataset\cite{8100027}. 

\subsection{Loss Function}
%% include loss function here
For fine-tuning the pre-trained Maskformer model, we used a main mask classification loss, $\mathcal{L}_{\text{mask-cls}}$, which is a linear combination of a cross-entropy classification loss for each predicted segmentation mask ($\mathcal{L}_{\text{CE}}$), a focal loss ($\mathcal{L}_{\text{focal}}$), and a dice loss ($\mathcal{L}_{\text{dice}}$), as shown below,
\begin{equation} 
\begin{aligned}
\mathcal{L}_{\text{mask-cls}} = \lambda_{\text{CE}} \mathcal{L}_{\text{CE}} + \lambda_{\text{focal}} \mathcal{L}_{\text{focal}} + \lambda_{\text{dice}} \mathcal{L}_{\text{dice}}
\end{aligned}
\end{equation}
The cross-entropy loss is used for learning the mask class while the focal loss and dice loss are used for learning the mask itself. The weights, $\lambda_{\text{CE}}$, $\lambda_{\text{focal}}$, and $\lambda_{\text{dice}}$ were empirically tuned using the validation set. In our experiments, we empirically found that $\lambda_{\text{focal}}=20$, $\lambda_{\text{dice}}=20$ and $\lambda_{\text{CE}}=1$ worked best for the dataset.

\subsection{Ensemble method}
Using an ensemble of models is a common strategy for achieving better performance in artificial intelligence and has also been used to improve the accuracy of models for the brain tumor segmentation task \cite{3D_ae_tumor}. We use an ensemble method as a way to combine the results of multiple MaskFormer models trained to segment GBM tumors using 2D axial, sagittal, and coronal slices in a 3D MRI volume (\autoref{fig:ensemble_example} (a-c)). We show in our results that an ensemble of models trained on all three orthogonal slice planes achieves superior performance than the traditional use of a single model trained only on 2D axial slices. In addition, for each slice direction, we independently train 3 different models with different strategies for a total of 9 models in our ensemble. 

At inference time, each model predicts segmentation masks in 2D slices in the same slicing direction as the training data. Next, we map the predictions for the 2D slices back to the 3D volume for all models. As a result, we obtain 9 sets of prediction volumes. We generate an ensemble segmentation mask by aggregating the predicted masks from all the nine models by majority voting for each voxel. 

% We can see that the predictions of the 3 models are consistent with the ground truth annotation with minor noticeable variations between the predictions \autoref{fig:ensemble_example} (d-g). 

\section{Dataset, Experiments and Results}

\subsection{Dataset}

We used UPenn-GBM, an openly available suite of preprocessed mpMRI scans curated by The University of Pennsylvania Health System hosted by the Cancer Imaging Archive\cite{UPennGBM, GBMDataRepo, TCIA}. The dataset consists of 3D MRI data for 611 unique subjects, each including four types of MRI imaging sequences: T1 (native T1-weighted), T1-GD (post-contrast T1), T2 (native T2-weighted) and FLAIR (T2 fluid-attenuated inversion recovery). Preprocessing includes skull-stripping and co-registration. Notably, the BraTS challenge dataset is a subset of this dataset and contains 173 samples\cite{UPennGBM}.

The dataset also contains segmentation masks labelling the various tumor regions and sub-regions: NCR (necrotic tumor core), ED (peritumoral edematous tissue), ET (enhancing tumor), and Else (everything else, i.e., healthy brain tissue). Of the 611 subjects, 147 contain segmentation masks annotated by experts which we use as ground truth labels during training \cite{UPennGBM}. Additionally, all 611 samples contain automatically generated segmentation masks using an ensemble of three previous BraTS challenge top-scoring models \cite{Kamnitsas_BraTS2015, Isensee_BraTS2020, McKinley_BraTS2020}.

\begin{figure}[t]
\begin{center}
\includegraphics[scale = 0.5]{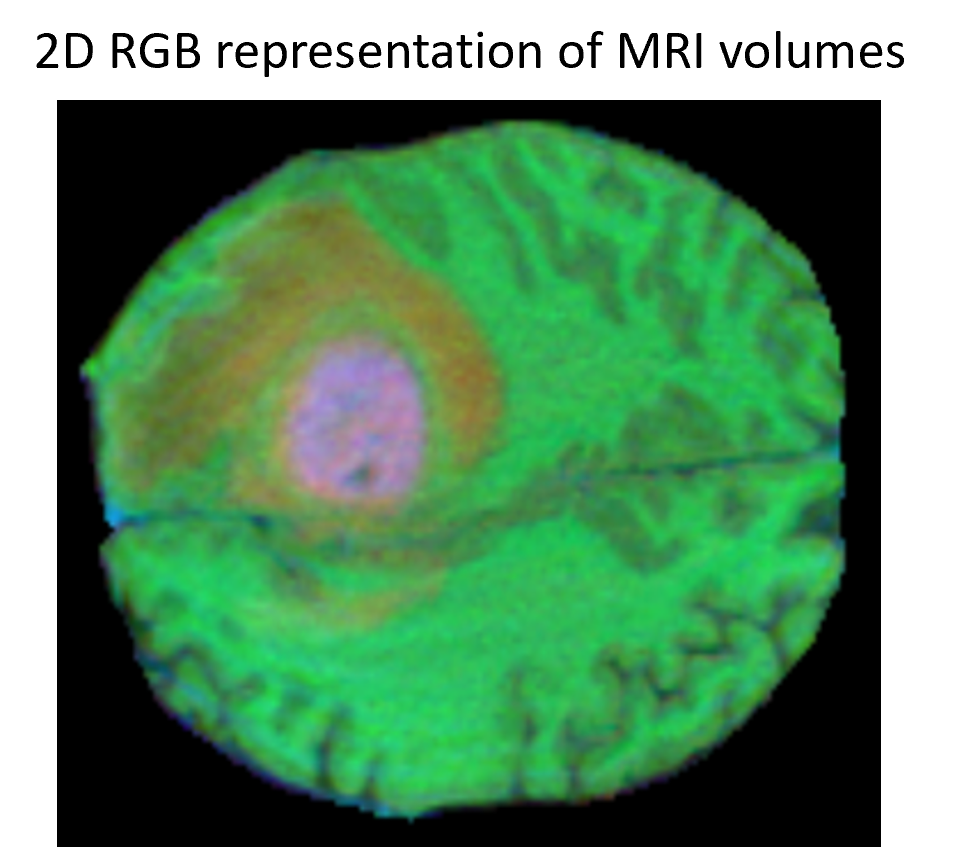}
\end{center}
\caption{An example of 2D RGB image after pre-processing the raw data.}
\label{fig:RGB_example}
\end{figure}

We preprocess the raw 3D MRI volumes into 2D RGB images using the following steps: 
\begin{enumerate}[label=(\roman*)]
    \item Data reduction: Remove zero slices across all volumes and obtain an image of size 163x193x146, i.e., 51.44\% of the original size.
    \item Data normalization: Scale each channel for each subject independently to be in [0, 1].
    \item 3D to 2D conversion: Convert the 3D volumes into 2D axial, sagittal, and coronal slices and keep track of their 3D volume positions.
    \item Data mapping: Select the FLAIR, T1, and T1-GD channels only and map them to RGB channels. Note that T2 is ignored because of the reported strong correlation between T1 and T2 channels for tumor segmentation\cite{UPennGBM}.
\end{enumerate}
An example of a preprocessed image slice is shown in \autoref{fig:RGB_example}.

\subsection{Experiments}

We made a 60:20:20 split for training, validation, and test data volumes respectively. The pre-trained MaskFormer model was finetuned using the training split and empirical design choices were made using the validation split. The following additional preprocessing steps were used: (i) normalize the images using the same mean and standard deviation as that used during the pre-training; (ii) resize the height and width of the input image. During the finetuning, we initialized the model with pre-trained model weights and allowed updates on all model parameters. We chose to update all model parameters rather than only the last several layers due to the different characteristics of the mpMRI data compared to the ADE20K dataset on which it was pretrained. We used the largest batch size possible with 15GB GPU memory. The actual batch size we used depended on the resolution of the input image and the size of the model (described below). The optimizer we used for training the model is the Adam optimizer.

To evaluate the model performance, we used the two main metrics in the BraTS challenge, the Dice Coefficient ($DC$) and 95\% Hausdorff distance ($HD_{95}$)\cite{BraTSchallenge}. Dice coefficient \cite{taha2015metrics} quantifies the similarity between a ground truth mask, $X$, and a predicted mask, $Y$, by calculating their overlap as,

\begin{equation}
\label{eq:dice}
DC = 2 \frac{|X \cap Y|}{|X| \cup |Y| }
\end{equation}

95\% Hausdorff distance \cite{reinke2022metrics, taha2015efficient} captures the spatial dissimilarity between two sets by calculating the maximum distance between any point in a reference set, $X$ to its closest point in the predicted set, $Y$, as,

\begin{equation}
\label{eq:hd}
HD_{95} (X, Y) = P_{95\%} 
\left\{ 
\sup_{x \in X} \inf_{y \in Y} \; d(x, y), \; \sup_{y \in Y} \inf_{x \in X} \; d(y, x)
\right\}
\end{equation}
where $P_{95\%}$ indicates the $95$-th percentile. Calculating the 95th percentile instead of the maximum value eliminates the impact of possible outliers.

We tested multiple different training schedules and empirically found the following two settings that performed the best: a constant learning rate of $10^{-5}$ and train for $20$ epochs; and, a cosine annealing learning rate scheduler\cite{DBLP:journals/corr/LoshchilovH16a} to vary learning rates throughout the training, starting with an initial value of $10^{-4}$ and train for $15$ epochs.

We implemented the following data augmentation strategies: 
\begin{itemize}
    \item Weak data augmentation\cite{Isensee_BraTS2021}: Jitter in HSV with probability $1$, horizontal flipping with probability $0.5$, and random cropping with a minimum $0.8$ of the image area with probability $0.8$.
    \item Strong data augmentation: Contrast (gamma) jitter with probability $0.5$, horizontal flipping with probability $0.3$, and random cropping with a minimum $0.8$ of the image area with probability $0.2$.
\end{itemize}

In our experiments, the small Maskformer model (63M parameters) outperformed both the base model (102M parameters) and the tiny model (42M parameters). We also experimented with the input image resolution (original resolution, $2\times$ original resolution, and $512\times512$ resolution). 

Finally, an ensemble method with the $3$ best combinations of model size, input resolution, data augmentation, and learning rate scheduler was used for the 2D axial, sagittal, and coronal slices individually for a total of $9$ models:
\begin{itemize}
\item model 1: small model; $2\times$ input resolution; strong data augmentation; fixed learning rate 
\item model 2: small model; $2\times$ input resolution; weak data augmentation; cosine annealing learning rate schedule 
\item model 3: small model; $512\times512$ input resolution; weak data augmentation; cosine annealing learning rate schedule
\end{itemize}

\subsection{Results}
\begin{figure}[t]
\centering
\subcaptionbox[]{Dice coefficients\label{fig:dice}}[.49\linewidth]{\includegraphics[trim=22 22 45 40, clip, scale=.3]{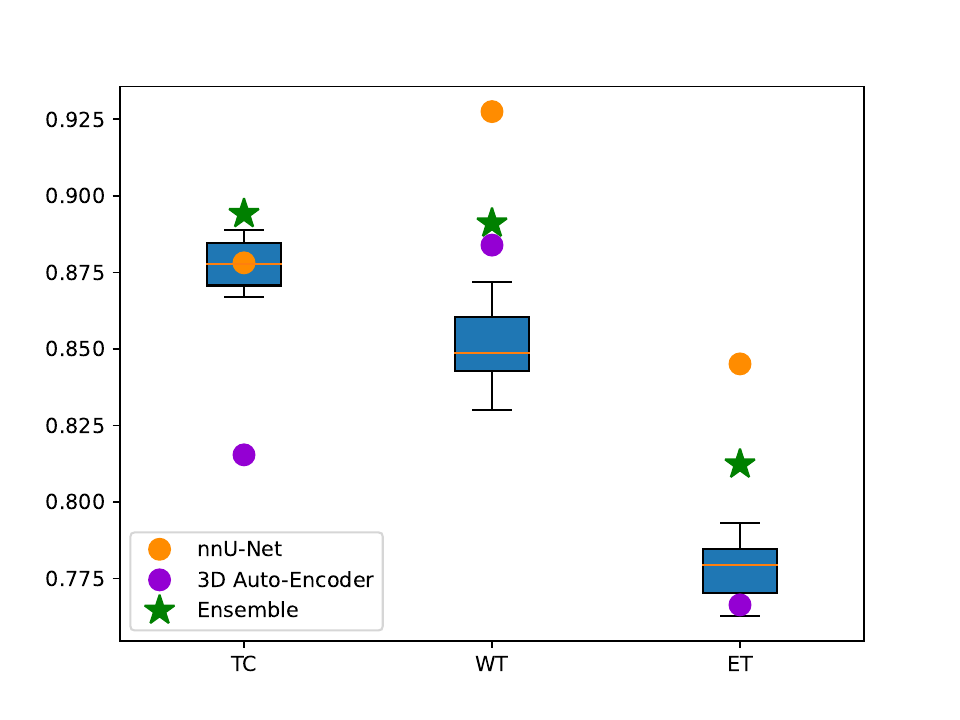}}
\subcaptionbox[]{95\% Hausdorff distance\label{fig:HD}}[.49\linewidth]{\includegraphics[trim=22 22 45 40, clip, scale=.3]{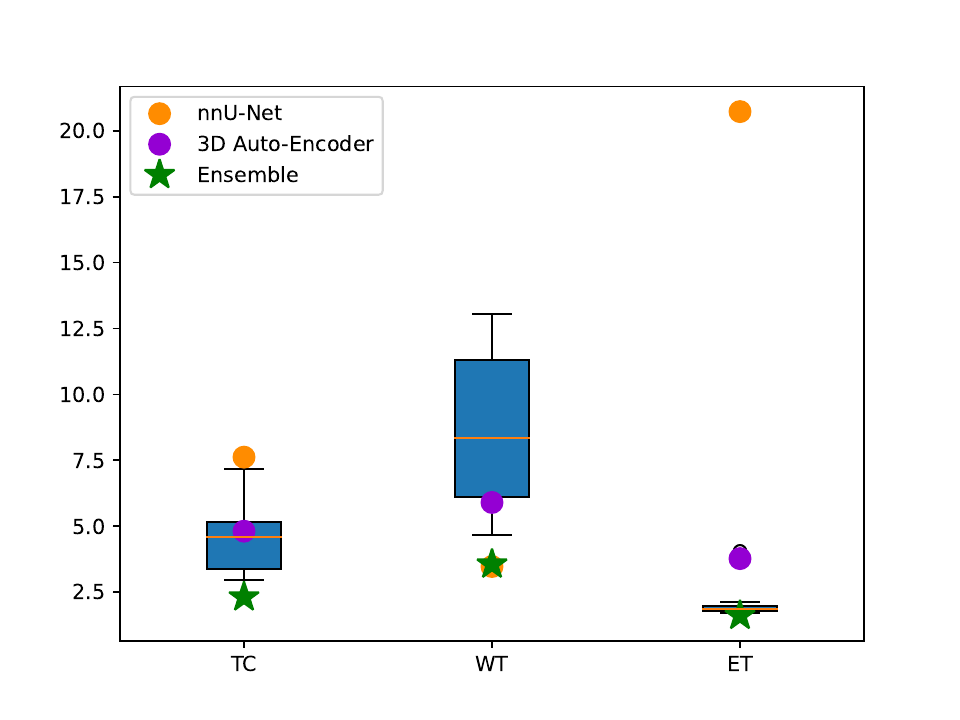}}
\caption{BRAINNET Model performance as compared with the performance of 3D autoencoder\cite{3D_ae_tumor} and nnU-Net\cite{Isensee_BraTS2020}. The blue boxes show the distribution of 9 individual models. Results are shown for tumor core (TC), whole tumor (WT), and enhancing tumor (ET) segmentations.}
\label{fig:benchmark_test}
\end{figure}

We trained BRAINNET to predict NCR, ED, and ET tumor segmentation masks. However, for the purposes of consistency while comparing with the state-of-the-art models, we report quantitative metrics (Dice coefficient, $DC$, and 95\% Hausdorff distance, $HD_{95}$) for the following:
\begin{itemize}
    \item Tumor core (TC): Combination of NCR and ET, i.e., $TC = NCR \; \cup \; ET$
    \item Whole tumor (WT): Combination of NCR, ET, and ED, i.e., $WT = NCR \; \cup \; ET \; \cup \; ED$
    \item Enhancing tumor (ET)
\end{itemize}

\begin{figure}[t]

    \begin{subfigure}[b]{0.45\textwidth} % Use textwidth for full width
        \centering
        \includegraphics[width=\textwidth]{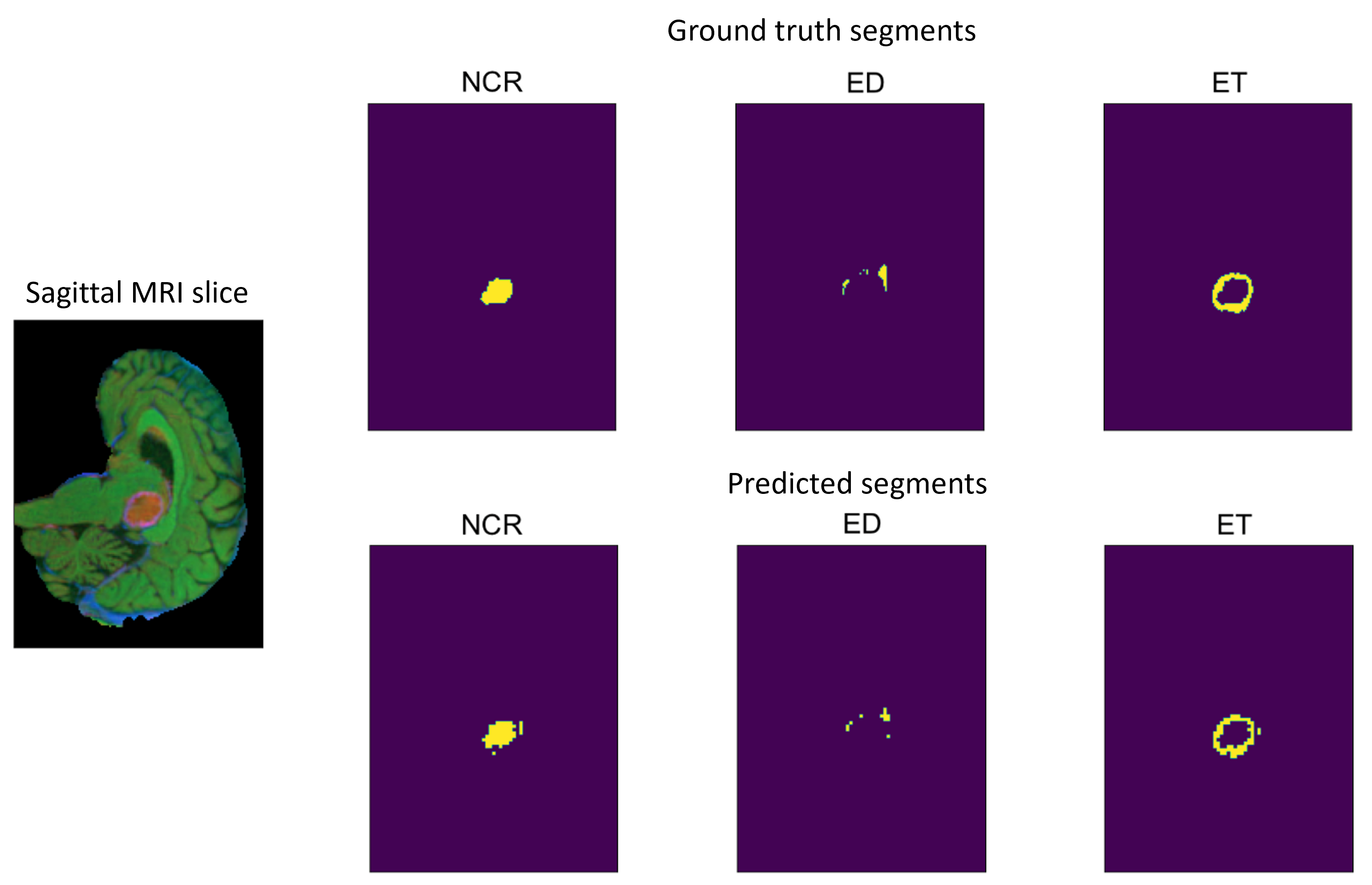}
    \end{subfigure}
    
    \caption{Examples of model predictions for three tumor classes (NCR, ED and ET).}
    \label{fig:prediction_example}
\end{figure}

We observed that the use of an ensemble resulted in a considerably better performance than each of the individual models(\autoref{fig:prediction_example}), resulting in higher $DC$ and lower $HD_{95}$. We also compared the performance of our BRAINNET model with two state-of-the-art models that were trained on the same dataset -- 3D autoencoder\cite{3D_ae_tumor} and nnU-Net\cite{Isensee_BraTS2020}. Comparing using $DC$, BRAINNET performed at par with these models and surpassed both these methods for TC segmentation. Using $HD_{95}$, BRAINNET outperformed both the models in predicting TC and ET while for WT, it performed at par. These results are also shown in \autoref{tab:results}.

\begin{table}[t]
    \caption{Performance comparison of BRAINNET with 3D autoencoder\cite{3D_ae_tumor} and nnU-Net\cite{Isensee_BraTS2020}. (The best performance for each tumor category using each performance metric is shown in bold.)}
	\begin{tabular}{| c || c | c | c || c | c | c |}
	\hline
    \multirow{ 2}{*}{Model} & \multicolumn{3}{c||}{$DC$} & \multicolumn{3}{c|}{$HD_{95}$} \\
    \cline{2-7}
	& TC & WT & ET & TC & WT & ET\\
    \hline
    \hline
    3D Autoencoder &0.815 &0.884 &0.766 &4.809 &4.516 &3.926 \\
    \hline
    nnU-Net &0.878 &\textbf{0.928} &\textbf{0.845} &7.623 &\textbf{3.47} &20.73 \\
    \hline
    BRAINNET & \textbf{0.894}& 0.891 &0.812 &\textbf{2.308} &3.552 &\textbf{1.608} \\
    \hline
	\end{tabular}
	\label{tab:results}
\end{table}

\begin{figure}[t]
\begin{center}
\includegraphics[scale = 0.4]{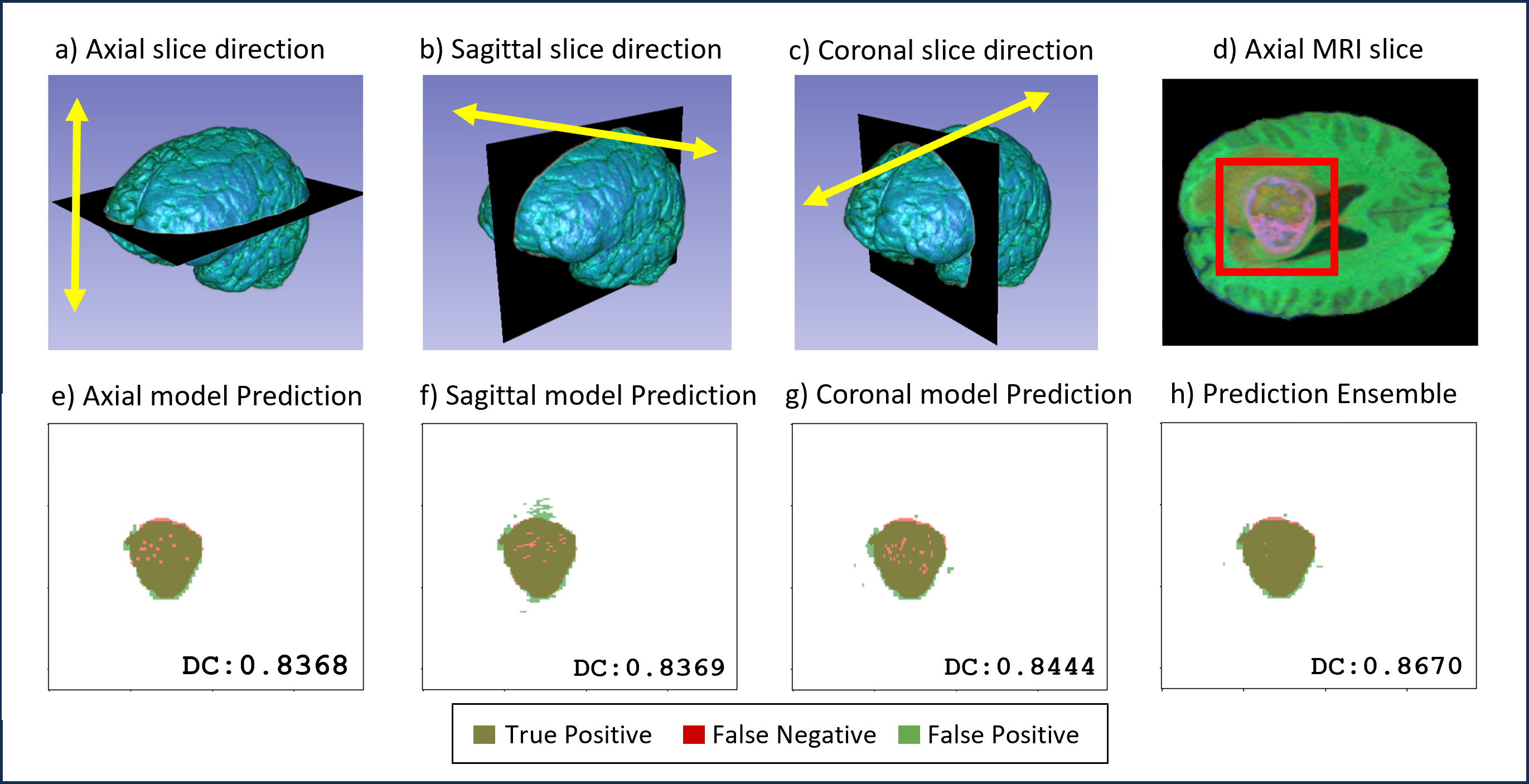}
\end{center}
\caption{Example of forming the prediction ensemble using individual model predictions: (a-c) MRI data slicing directions, (d) an example of input MRI slice in axial direction, (e-g) comparison of model predictions at the example slice with the ground truth annotation, (h) comparison of ensemble prediction with ground truth annotation. Note how the false positive and false negative predictions in the individual slice directions are eliminated in the ensemble prediction.}
\label{fig:ensemble_example}
\end{figure}

An example of BRAINNET ensemble model predictions alongside the ground truth for all 3 tumor segmentation categories along with the corresponding sagittal MRI slice is shown in \autoref{fig:prediction_example}. We observed that false positive and false negative tumor predictions for each model in each slice direction are ultimately eliminated to a large extent by using the ensemble approach (an example is shown in \autoref{fig:ensemble_example}). For simplicity, \autoref{fig:ensemble_example} only shows the prediction of one model in each slice direction.

\section{Discussion and Conclusion}

Automated brain tumor segmentation is targeted to help alleviate the workload of radiologists, who would have to now review segmentation masks instead of following the time-intensive process of creating the masks themselves. In this paper, we have introduced BRAINNET, a pipeline for segmenting different GBM tumor regions from mpMRI images using an ensemble of fine-tuned MaskFormer models that combines predictions of nine models, three along each 2D orthogonal slice direction (axial, sagittal, and coronal). Experiments showed that BRAINNET achieves state-of-the-art performance in predicting all the tumor regions using only 15GB GPU memory. The model was trained and tested on the publicly available UPenn-GBM dataset, from which the BraTS challenge uses a subset. Our model achieved performance that is comparable with or surpasses other state-of-the-art models when evaluated using the Dice coefficient and 95\% Hausdorff distance. In the future, we will further investigate the possibility of improving performance using more sophisticated ensemble methods that leverage the raw voxel-wise probability outputs of the individual segmentation models instead of the segmentation masks.  

The source code for BRAINNET is publicly available at \url{https://github.com/RiceD2KLab/BRAINNET}.

\section{Compliance with Ethical Standards}
This study was retrospectively conducted using human subject data that have been made openly available by The University of Pennsylvania Health System and hosted by the Cancer Imaging Archive \cite{UPennGBM, GBMDataRepo, TCIA}.
Ethical approval was not required for this study as confirmed by the open access license (Creative Commons Attribution 3.0 Unported License).

\section{Conflicts of Interest}
The authors have no relevant financial or non-financial interests to disclose. No funding was received for this study.

\bibliographystyle{IEEEbib}

\counterwithin{figure}{section}
\counterwithin{table}{section}

\end{document}